\documentclass[12pt]{article}

\newcommand{\be}{\begin{equation}}
\newcommand{\ee}{\end{equation}}
\newcommand{\bi}[1]{\vspace{-3mm} \bibitem{#1}}

\begin{document}

\begin{center}

{\it International Journal of Modern Physics B 19 (2005) 4103-4114} 

\vskip 5 mm
{\Large \bf Dynamics of Fractal Solids}
\vskip 5 mm

{\large \bf Vasily E. Tarasov } \\

\vskip 3mm
{\it Skobeltsyn Institute of Nuclear Physics, \\
Moscow State University, Moscow 119992, Russia}

{E-mail: tarasov@theory.sinp.msu.ru}
\end{center}
\vskip 11 mm

\begin{abstract}
We describe the fractal solid by a special continuous medium model. 
We propose to describe the fractal solid by a fractional continuous model, 
where all characteristics and fields are defined 
everywhere in the volume but they follow some generalized 
equations which are derived by using integrals of fractional order.
The order of fractional integral can be equal to 
the fractal mass dimension of the solid.
Fractional integrals are considered as an approximation 
of integrals on fractals.
We suggest the approach to compute the moments of inertia 
for fractal solids. 
The dynamics of fractal solids are described by the usual Euler's
equations.
The possible experimental test of 
the continuous medium model for fractal solids is considered. 

\end{abstract}

\vskip 3 mm


%03.40.-t	Classical mechanics of continuous media: 
%%%general mathematical aspects
%05.45.Df Fractals
%47.53.+n	Fractals

\section{Introduction}

Derivatives and integrals of fractional order \cite{SKM,OS,MR}
have found many applications in recent studies in condensed matter physics.
The interest to fractional analysis has grown continually in the last years. 
Fractional analysis has numerous applications: kinetic theories 
\cite{Zaslavsky1,Zaslavsky2,Zaslavsky3,Zaslavsky4,PhysicaA,Chaos2005},   
statistical mechanics of fractal systems \cite{Chaos2004,PRE2005,JPCS},  
dynamics in a complex or 
porous media \cite{Nig,PLA2005-1,AP2005-2,MPLB2005-1,PLA2005-2,CMDA}, 
electrodynamics \cite{En1,En2,En3,En4,POP,MPLB2005-2}, 
and many others. 

In order to use fractional derivatives and fractional integrals 
for fractal media, we must use some continuous medium model \cite{PLA2005-1}. 
We propose to  describe the fractal medium by a fractional continuous
medium \cite{PLA2005-1,AP2005-2}, 
where all characteristics and fields are defined 
everywhere in the volume but they follow some generalized 
equations which are derived by using fractional integrals.
In many problems the real fractal structure of matter
can be disregarded and the medium can be replaced by
some fractional continuous mathematical model.
Smoothing of the microscopic characteristics over the
physically infinitesimal volume transforms the initial
fractal medium into fractional continuous model \cite{PLA2005-1}
that uses the fractional integrals.
The order of fractional integral is equal
to the fractal mass dimension of the medium.
The fractional integrals allow us to take into account the
fractality of the media.
In order to describe the fractal medium by continuous medium model
we must use the fractional integrals, which 
are considered as an approximation of integrals on fractals. 
In Ref. \cite{RLWQ}, authors prove that integrals
on net of fractals can be approximated by fractional integrals.
In Ref. \cite{Chaos2004,PRE2005,JPCS}, we proved that fractional integrals
can be considered as integrals over the space with fractional
dimension up to numerical factor. 
To prove this, we use the formulas of dimensional regularization \cite{Col}.

In this paper,
we use the fractional integrals to describe fractal solids.
We consider the fractal solid by using the fractional continuous medium model. 
To describe a fractal  solid, we use integrals of fractional order. 
We prove that equations of motion for fractal solid 
have the same form as the equations for usual solid.
We suggest the approach to compute the moments of inertia 
for fractal solids, and consider the possible experimental testing 
of the continuous medium model for fractal solid.

%%%%%%%%%%%%%%%%%%%%%%%%%%%%%%%%%%%%%%%%%%%%%%%%%%%%%%%%%%%%%%%%%%%%%%%%%%
\section{Mass of Fractal Solid}

The fractal solid is characterized by the fractal dimensions. 
It is known that fractal dimension can be best calculated by 
box-counting method which means drawing a box of size $R$ 
and counting the mass inside. 
The mass fractal dimension \cite{Mand,Feder} can be easily measured
for fractal solids.
The properties of the fractal solid like mass obeys a power law 
relation $M \sim R^{D}$, 
where $M$ is the mass of fractal solid, $R$ 
is a box size (or a sphere radius),
and $D$ is a mass fractal dimension. 
The power law relation $M \sim R^{D}$ can be naturally 
derived by using the fractional integral.
In Ref. \cite{PLA2005-1}, we prove that the mass fractal dimension 
is connected to the order of fractional integrals. 

Let us consider the region $W$ of solid 
in three-dimensional Euclidean space $E^3$.
The volume of the region $W$ is denoted by $V(W)$.
The mass of the region $W$ in the fractal solid is denoted by $M(W)$. 
The fractality of solid means that the mass of this solid 
in any region $W$ of Euclidean space $E^3$ increases more slowly 
that the volume of this region.
For the ball region of the fractal solid, 
this property can be described by the power law $M \sim R^{D}$, 
where $R$ is the radius of the ball $W$. 

Fractal solid is called a homogeneous if 
the following property is satisfied:
for all regions $W$ and $W^{\prime}$ of the homogeneous fractal solid 
such that the volumes are equal $V(W)=V(W^{\prime})$, 
we have the masses of these regions equal too, i.e., 
$M(W)=M(W^{\prime})$. 
Note that the wide class of the fractal media satisfies 
the homogeneous property.
In Ref. \cite{PLA2005-1,AP2005-2}, the fractional continuous medium 
model for the fractal media has been suggested. 
Note that the fractality and homogeneity properties 
in the fractional continuous model are 
realized in the following forms: 

\noindent
(1) Homogeneity: \\
The local density of the homogeneous fractal solid in 
the continuous model has the form $\rho({\bf r})=\rho_0=const$.

\noindent
(2) Fractality:
The mass of the ball region $W$ of fractal solid obeys a power law relation
$M \sim R^{D}$, where $D<3$, and $R$ is the radius of the ball. \\

The mass of the region $W$ in the solid with integer mass dimension
is derived by the equation
realized by the 
fractional generalization of the equation
\be \label{MW} M_3(W)=\int_W \rho({\bf r}) d V_3  . \ee
We ca consider the fractional generalization of this equation.
Let us define the fractional integral
in Euclidean space $E^3$ in the Riesz form \cite{SKM}.
The fractional generalization of Eq. (\ref{MW}) can be realized
in the following form
\be \label{ID} M_D(W)=\int_W \rho({\bf r}) dV_D , \ee
where $dV_D=c_3(D,{\bf r})d V_3$, and 
\be \label{IDc}
c_3(D,{\bf r})=\frac{2^{3-D} \Gamma(3/2)}{\Gamma(D/2)} |{\bf r}|^{D-3} . \ee
Here, we use the initial points in the fractional integrals which are set to zero.
The numerical factor in Eqs. (\ref{ID}) and (\ref{IDc}) 
has this form in order to derive 
the usual integral in the limit $D\rightarrow (3-0)$.
Note that the usual numerical factor
$\gamma^{-1}_3(D)={\Gamma(1/2)}/{2^D \pi^{3/2} \Gamma(D/2)}$,
which is used in Ref. \cite{SKM} 
leads to $\gamma^{-1}_3(3-0)= {\Gamma(1/2)}/{2^3 \pi^{3/2} \Gamma(3/2)}
=1/(4\pi^{3/2})$ in the limit $D\rightarrow (3-0)$. 

In order to have the usual dimensions of the physical values,
we can use vector ${\bf r}$, and coordinates 
$x$, $y$, $z$ as dimensionless values.

We can rewrite Eq. (\ref{ID}) in the form
\be \label{MWD}  M_D(W)=\frac{2^{3-D} \Gamma(3/2)}{\Gamma(D/2)}
\int_W \rho({\bf r}) |{\bf r}|^{D-3} d V_3 . \ee

If we consider the homogeneous fractal solid 
($\rho({\bf r})=\rho_0=const$) and the ball region 
$W=\{{\bf r}: \  |{\bf r}|\le R \}$, then we have 
\be M_D(W)= \rho_0 \frac{2^{3-D} \Gamma(3/2)}{\Gamma(D/2)} 
\int_W |{\bf r}|^{D-3} d V_3 . \ee
Using the spherical coordinates, we get
\be M_D(W)= \frac{\pi 2^{5-D} \Gamma(3/2)}{\Gamma(D/2)} \rho_0 
\int_W |{\bf r}|^{D-1} d |{\bf r}|= 
\frac{2^{5-D} \pi \Gamma(3/2)}{D \Gamma(D/2)} \rho_0 R^{D} . \ee
As the result, we have $M(W)\sim R^D$, i.e., we derive 
equation $M \sim R^{D}$ up to the numerical factor.
Therefore the fractal solid with non-integer mass dimension $D$ can be
described by fractional integral of order $D$.
Note that the interpretation of the fractional integration
is connected with fractional dimension \cite{Chaos2004,PRE2005}.
This interpretation follows from
the well-known formulas for dimensional regularizations \cite{Col}.
The fractional integral can be considered as a 
integral in the fractional dimension space up to the numerical 
factor $\Gamma(D/2) /( 2 \pi^{D/2} \Gamma(D))$.

%%%%%%%%%%%%%%%%%%%%%%%%%%%%%%%%%%%%%%%%%%%%%%%%%%%
\section{Moment of Inertia of Fractal Solid}

\subsection{Fractional equation for moment of inertia}

%%%In this section, we consider the fractional generalization of equations
%%%for moments of inertia.

The moment of inertia of a solid body with density $\rho({\bf r})$ 
with respect to a given axis is defined by the volume integral
\be \label{(1)}
I=\int_W \rho({\bf r}) \, {\bf r}^2_{\perp} \, dV_3,
\ee
where ${\bf r}^2_{\perp}$ is the perpendicular distance from 
the axis of rotation. 
This can be broken into components as
\be \label{(3)}
I_{kl}=\int_W \rho({\bf r}) \, ({\bf r}^2 \delta_{kl}-x_kx_l) \, dV_3
\ee
for a continuous mass distribution. 
Here, ${\bf r}=x_k{\bf e}_k$ is 
the distance to a point (not the perpendicular distance) 
and $\delta_{kl}$ is the Kronecker delta. Depending on the context, 
$I_{kl}$ may be viewed either as a tensor or a matrix. 

The fractional generalization of equation (\ref{(3)}) has the form
\be \label{(3F)}
I^{(D)}_{kl}=\int_W \rho({\bf r})({\bf r}^2 \delta_{kl}-x_kx_l) dV_D,
\ee
where $dV_D=c_3(D,{\bf r}) dV_3$.
The moment of inertia tensor is symmetric ($I^{(D)}_{kl}=I^{(D)}_{lk}$).
%%%and is related to the angular momentum vector 
%%%${\bf L}$ by $L_k=I^{(D)}_{kl}\omega_l$, 
%%%where ${\bf \omega}=\omega_k {\bf e}_k$ is the angular velocity vector.

The principal moments are given by the entries 
in the diagonalized moment of inertia matrix. 
The principal axes of a rotating body are defined 
by finding values of $\lambda$ such that
\be (I^{(D)}_{kl}-\lambda \delta_{kl}) \omega_{l}=0 , \ee
which is an eigenvalue problem.
Here, ${\bf \omega}=\omega_k {\bf e}_k$ is the angular velocity vector.
The tensor $I^{(D)}_{kl}$ may be diagonalized 
by transforming to appropriate coordinate system. 
The moments of inertia in the coordinate system, 
corresponding to the eigenvalues of the tensor, 
are known as principal moments of inertia.

\subsection{Moment of inertia of fractal solid sphere}

For a fractal solid sphere  with radius R, and mass M, 
the moment of inertia can be derived by Eq. (\ref{(3F)}).
The moment of inertia can be computed directly by noting 
that the component of the radius perpendicular 
to the z-axis in spherical coordinates  is
\be {\bf r}^2_{\perp} =(r \ sin \phi)^2, \ee
where $\phi$ is the angle from the z-axis.
Using the fractional generalization of Eq. (\ref{(1)}), we have 
\[ I^{(D)}_z=\int_W \rho({\bf r}) {\bf r}^2_{\perp} dV_D= \]
\[ =\frac{2^{3-D} \Gamma(3/2)}{\Gamma(D/2)}
\int^R_0 \int^{2\pi}_0 \int^{\pi}_0 \rho({\bf r}) 
(r \ sin \phi)^2 r^{D-1} sin \phi \ d \phi d \theta dr =\]
\[ =\frac{2^{3-D} \Gamma(3/2)}{\Gamma(D/2)}
\int^R_0 \int^{2\pi}_0 \int^{\pi}_0 \rho({\bf r}) 
r^{D+1} \ sin^3 \phi \ d \phi d \theta dr = \]
\[ =\frac{2^{3-D} \Gamma(3/2)}{\Gamma(D/2)}
\int^R_0 \int^{2\pi}_0 \int^{\pi}_0 \rho({\bf r}) 
r^{D+1} \ (1-cos^2 \phi) sin \phi \ d \phi d \theta dr . \]
Making the change of variables
\be \label{u} u=cos \phi , \quad du=-sin \phi \ d\phi , \ee
we then allow the integral to be written simply and solved by quadrature.
For homogeneous fractal solid sphere ($\rho({\bf r})=\rho_0$), we have
\[ I^{(D)}_z=\frac{2^{3-D} \Gamma(3/2)}{\Gamma(D/2)}
\int^R_0 \int^{2\pi}_0 \int^1_{-1} \rho_0 
r^{D+1} \ (1-u^2) d u d \theta dr = \]
%%%\[ =\frac{2^{3-D} \Gamma(3/2)}{\Gamma(D/2)}
%%%\int^R_0 \int^{2\pi}_0  \rho_0  
%%%r^{D+1} \ (u-\frac{1}{3}u^3)^{+1}_{-1} d \theta dr = \]
\[ =\frac{2^{5-D} \Gamma(3/2)}{3\Gamma(D/2)}
\int^R_0 \int^{2\pi}_0 \rho_0 r^{D+1}  d \theta dr = 
\frac{\pi 2^{6-D} \Gamma(3/2)}{3\Gamma(D/2)} \rho_0
\int^R_0 r^{D+1} dr .  \]
As the result, we get
\be \label{IS}
I^{(D)}_z=
\frac{\pi 2^{6-D} \Gamma(3/2)}{3(D+2)\Gamma(D/2)} \rho_0 R^{D+2} . \ee

The mass of the fractal solid sphere is defined by Eq. (\ref{ID}). 
Therefore, we have
\[ M_D=\int_W \rho({\bf r}) dV_D=
\frac{2^{3-D} \Gamma(3/2)}{\Gamma(D/2)}
\int^R_0 \int^{2\pi}_0 \int^{\pi}_0 \rho({\bf r}) 
r^{D-1} sin \phi \ d \phi d \theta dr =\]
\[ =\frac{2^{3-D} \Gamma(3/2)}{\Gamma(D/2)}
\int^R_0 \int^{2\pi}_0 \int^{\pi}_0 \rho({\bf r}) 
r^{D-1} \ sin \phi \ d \phi d \theta dr . \]
Making the change of variables (\ref{u})
then allows the integral to be written simply and solved by quadrature.
For homogeneous solid sphere ($\rho({\bf r})=\rho_0$), we get
\[ M_{D}=\frac{2^{3-D} \Gamma(3/2)}{\Gamma(D/2)}
\int^R_0 \int^{2\pi}_0 \int^1_{-1} \rho_0 
r^{D+1} \ d u d \theta dr = \]
\[ =\frac{2^{3-D} \Gamma(3/2)}{\Gamma(D/2)}
\int^R_0 \int^{2\pi}_0  \rho_0 
r^{D+1} \ (u)^{+1}_{-1} d \theta dr = \]
\[ =\frac{2^{4-D} \Gamma(3/2)}{\Gamma(D/2)}
\int^R_0 \int^{2\pi}_0  \rho_0 r^{D-1}  d \theta dr = 
\frac{\pi 2^{5-D} \Gamma(3/2)}{\Gamma(D/2)} \rho_0 
\int^R_0 r^{D-1}  dr . \]
As the result, we have
\be \label{MS}
M_D=\frac{\pi 2^{5-D} \Gamma(3/2)}{D \Gamma(D/2)} \rho_0 R^{D} . \ee

Substituting $\rho_0$ from Eq. (\ref{MS}) in Eq. (\ref{IS}), 
we get the moment of inertia for fractal solid sphere in the form
\be \label{IS2}
I^{(D)}_z=\frac{2D}{3(D+2)} M_D R^2 .
\ee
If $D=3$, then we have the usual relation $I^{(3)}_z=(2/5)MR^2$.
If $D=(2+0)$, then we have 
$I^{(2+0)}_z=(1/3)MR^2$. Note that fractal solid sphere
with dimension $D=(2+0)$ cannot be considered as a 
spherical shell that has $I_z=(2/3)MR^2$. 
In fractal solid sphere, we have the
homogeneous distribution of fractal matter in the volume. 

Because of the symmetry of the sphere, each principal moment is the same, 
so the moment of inertia of the sphere taken about 
any diameter is Eq. (\ref{IS2}). 

The moments of inertia $I^{(D)}_z$ and $I^{(3)}_z$ are connected 
by the relation
\be I^{(D)}_z/I^{(3)}_z=1+\frac{2(D-3)}{3(D+2)}  . \ee
Using $2<D\le 3$, we get $(5/6)<I^{(D)}_z/I^{(3)}_z\le 1$.

\subsection{Moment of inertia for fractal solid cylinder}

The equation for the moment of inertia of homogeneous cylinder with
integer mass dimension has the well-known form
\be \label{0} 
I^{(2)}_z=\rho_0 \int _S (x^2+y^2) dS_{2} \int_L dz . \ee
Here $z$ is the cylinder axis, and $dS_2=dxdy$. 
The fractional generalization of Eq. (\ref{0}) 
can be defined by the equation
\be \label{1} 
I^{(\alpha)}_z=\rho_0 \int _S (x^2+y^2) dS_{\alpha} \int_L dl_{\beta} , \ee
where we use the following notations
\be \label{2} dS_{\alpha}=c(\alpha) (\sqrt{x^2+y^2})^{\alpha-2} dS_2 , 
\quad dS_2=dxdy , 
\quad c(\alpha)=\frac{2^{2-\alpha}}{\Gamma(\alpha/2)} , \quad
dl_{\beta}= \frac{|z|^{\beta-1}}{\Gamma(\beta)} dz . \ee
The numerical factor in Eq. (\ref{1}) has this form in order to
derive usual integral in the limit $\alpha\rightarrow (2-0)$ and 
$\beta \rightarrow (1-0)$.
The parameters $\alpha$ and $\beta$ are
\[ 1 < \alpha \le 2, \quad 0 < \beta \le 1 . \]
If $\alpha=2$ and $\beta=1$, then Eq. (\ref{1}) has form (\ref{0}).
The parameter $\alpha$ is a fractal mass dimension of the cross-section 
of the cylinder. This parameter can be easy calculated from
the experimental data. It can be calculated by box-counting method 
for the cross-section of the cylinder.

Substituting Eq. (\ref{2}) in Eq. (\ref{1}), we get
\be I^{(\alpha)}_z=
\frac{\rho_0 c(\alpha)}{\Gamma(\beta)} \int_S (x^2+y^2)^{\alpha/2} dS_2 
\int^H_0 z^{\beta-1} dz . \ee
Here, we consider the cylindrical region $W$ that is defined by the relations
\be \label{cyl}  L=\{z: \ 0\le z\le H \}, 
\quad S=\{(x,y): \ 0\le x^2+y^2 \le R^2 \} . \ee
Using the cylindrical coordinates $(\phi,r,z)$, we have
\be dS_2=dxdy=r dr d\phi , \quad (x^2+y^2)^{\alpha/2}=r^{\alpha} . \ee
Therefore the moment of inertia is defined by
\be I^{(\alpha)}_z=\frac{2 \pi \rho_0 c(\alpha)}{\Gamma(\beta)} 
\int^R_0 r^{\alpha+1} dr \int^H_0 z^{\beta-1} dz=  
\frac{2 \pi \rho_0 c(\alpha)}{(\alpha+2) \beta \Gamma(\beta)} 
R^{\alpha+2} H^{\beta} . \ee
As the result, we have the moment of inertia of the fractal solid
cylinder in the form 
\be \label{11} I^{(\alpha)}_z=
\frac{2 \pi \rho_0 c(\alpha)}{(\alpha+2) \beta 
\Gamma(\beta)} R^{\alpha+2} H^{\beta} . \ee
If $\alpha=2$ and $\beta=1$, we get $I^{(2)}_z=(1/2)\pi\rho_0 R^4H$.

The mass of the usual homogeneous 
cylinder (\ref{cyl}) is defined by the equation
\be \label{M2} M=\rho_0 \int_S dS_{2} \int_L dz= 
2 \pi \rho_0  \int^R_0 r dr \int^H_0 dz=  \pi \rho_0 R^2 H . \ee
We can consider the fractional generalization of this equation.
The mass of the fractal solid cylinder (\ref{cyl}) can be defined 
by the equation
\be \label{Ma} M_{\alpha}=\rho_0 \int_S dS_{\alpha} \int_L dl_{\beta} , \ee
where $dS_{\alpha}$ and $dl_{\beta}$ are defined by Eq. (\ref{2}). 
Using the cylindrical coordinates, we get the mass of fractal 
solid cylinder in the form 
\be M_{\alpha}=\frac{2 \pi \rho_0 c(\alpha)}{\Gamma(\beta)} 
\int^R_0 r^{\alpha-1} dr 
\int^H_0 z^{\beta-1} dz=  
\frac{2 \pi \rho_0 c(\alpha)}{\alpha \beta \Gamma(\beta)} 
R^{\alpha} H^{\beta} . \ee
As the result, we have
\be \label{M-F} 
M_{\alpha}=\frac{2 \pi \rho_0 c(\alpha)}{\alpha \beta \Gamma(\beta)} 
R^{\alpha} H^{\beta} . \ee

Substituting mass (\ref{M-F}) in the moment of inertia (\ref{11}), 
we get the relation
\be \label{I-F} I^{(\alpha)}_z=\frac{\alpha}{\alpha+2} M_{\alpha} R^2 . \ee
Note that Eq. (\ref{I-F}) has not the parameter $\beta$. 
If $\alpha=2$, we have the well-known relation $I^{(2)}_z=(1/2) M R^2$ 
for the homogeneous cylinder that has the integer mass dimension $D=3$
and $\alpha=2$.

Let us consider the fractal solid cylinder with the mass and radius 
that are equal to mass and radius of 
the homogeneous solid cylinder with integer mass dimension.
In this case, the moments of inertia of these cylinders 
are connected by the equation
\be I^{(\alpha)}_z=\frac{2\alpha}{\alpha+2} I^{(2)}_z .\ee
Here, $I^{(2)}_z$ is the moment of inertia for the
cylinder with integer mass dimension $D=3$ and $\alpha=2$.
For example, the parameter $\alpha=1.5$ leads us to the relation
$I^{(3/2)}=(6/7) I^{(2)}_z$.
Using $1\le \alpha \le 2$, we have the relation
\be (2/3) \le I^{(\alpha)}_z/ I^{(2)}_z \le 1 . \ee

As the result, 
the fractal solid cylinder with the mass $M$, and radius $R$,
has the moment of inertia $I^{(\alpha)}_z$ such that
\be I^{(\alpha)}_z/I^{(2)}_z=1+\frac{\alpha-2}{\alpha+2} , \ee
where $\alpha$ is a fractal mass dimension of cross-section of
the cylinder ($1<\alpha \le 2$). The parameter $\alpha$ can be calculated 
by box-counting method for the cross-section of the cylinder.
Here, $I^{(2)}$ is the moment of inertia of usual cylinder 
with the mass $M$, and radius $R$.

%%%%%%%%%%%%%%%%%%%%%%%%%%%%%%%%%%%%%%%%%%%%%%%%%%%%%%%%%%%%%%%%%%%%%%%%%%
\section{Equations of Motion for Fractal Solid}

\subsection{Euler's equations for fractal solid}

The moment of momentum ${\bf L}=L_k {\bf e}_k$ is
defined by the equation
\be \label{L3} {\bf L}=\int_W [{\bf r},{\bf v}] \rho({\bf r}) \ dV_3, \ee
where $[\ , \ ]$ is a vector product.
The vector ${\bf r}=x_k {\bf e}_k$ is a radius vector, and 
${\bf v}=v_k {\bf e}_k$ is a velocity of points with masses 
$dM_3=\rho({\bf r})dV_3$.
The fractional generalization of Eq. (\ref{L3}) has the form
\be \label{L} {\bf L}^{(D)}=\int_W [{\bf r},{\bf v}] \rho({\bf r}) \ dV_D, \ee
where we use $dM_D= \rho({\bf r})dV_D$.
Using ${\bf v}=[{\bf \omega},{\bf r}]$, we get moment of momentum 
in the form
\be \label{LI} L^{(D)}_k=I^{(D)}_{kl}\omega_{l} . \ee
The moment of inertia tensor $I^{(D)}_{kl}$ is  
related to the angular momentum vector 
${\bf L}^{(D)}$ by Eq. (\ref{LI}), 
where ${\bf \omega}=\omega_k {\bf e}_k$ is the angular velocity vector.

For a fractal solid with one point fixed, if the angular momentum 
${\bf L}^{(D)}$ is measured in the frame of the rotating body, we have 
the equation 
%%%(as shown in the equations for torque),
\be \label{bfL}
\frac{d{\bf L^{(D)}}}{dt}+[{\bf \omega},{\bf L}^{(D)}]={\bf N},
\ee
where ${\bf \omega}$ is the angular velocity vector and 
${\bf N}=N_k {\bf e}_k$ is the torque (moment of force).
%%%where ${\bf N}$ is the moment of force 
%%%about point O.
For components, we have
\be \label{dL}
\frac{dL^{(D)}_k}{dt}+\varepsilon_{klm} \omega_l L^{(D)}_m=N_k ,
\ee
where $L_k$ are defined by the relations
\be L^{(D)}_k=\int_W \rho({\bf r}) \varepsilon_{klm} x_l v_m dV_D .\ee
Here, $\varepsilon_{klm}$ is the permutation symbol, 
${\bf \omega}=\omega_k {\bf e}_k$ is the angular frequency, 
and ${\bf N}=N_k {\bf e}_k$ is the external torque.

If the principle body axes are chosen, $L^{(D)}_k=I^{(D)}_k \omega_k$, then
\be \label{Io}
\frac{d(I^{(D)}_k \omega_k)}{dt}+
\varepsilon_{klm} \omega_l \omega_m I^{(D)}_m=N_k .
\ee
These are Euler's equations of motion. 
Taking the principal axes frame, we get
the Euler's equations of motion for fractal solid in the form
\[ I^{(D)}_x \frac{d\omega_x}{dt}+
(I^{(D)}_z-I^{(D)}_y)\omega_y \omega_z = N_x, \]
\[ I^{(D)}_y \frac{d\omega_y}{dt}+
(I^{(D)}_x-I^{(D)}_z)\omega_x \omega_z = N_y, \]
\[ I^{(D)}_z \frac{d\omega_z}{dt}+
(I^{(D)}_y-I^{(D)}_x)\omega_x \omega_y = N_z, \]
where $I^{(D)}_x$, $I^{(D)}_y$, and $I^{(D)}_z$ 
are the principal moments of inertia.
As the result, we proved that equations of motion for fractal solid 
have the same form as the equations for usual solids.

For general non-rigid motion, the equation of motion is Liouville's 
equation \cite{Chaos2004,PRE2005}, which 
can be considered as the generalization of 
Euler's equations of motion to systems that are not rigid. 
In Eulerian form, the rotating axes are chosen to coincide 
with the instantaneous principle axes of the continuous system. 
For general non-rigid motion, Euler's equations are then replaced 
by Eq. (\ref{bfL}) or, in component form (\ref{dL}). 
The extension of the Liouville equation to include 
collisions is known as the Bogoliubov equations \cite{PRE2005,JPCS}.

%%%%%%%%%%%%%%%%%%%%%%%%%%%%%%%%%%%%%%%%%%%%%%%%%%
\subsection{Pendulum with fractal solid}

In this section, we consider the possible experimental testing 
of the continuous medium model for fractal solid. 
In this test we suggest measuring the period of 
pendulum with a fractal solid which is unequal to the 
period of the usual solid with the same mass and form. 

Let us consider the Maxwell pendulum with fractal solid cylinder.
Usually, the Maxwell pendulum is used to demonstrate transformations 
between gravitational potential energy and rotational kinetic energy.
The device has some initial gravitational potential energy,
when the string is winding on the small axis. 
When released, this gravitational 
potential energy is converted into rotational kinetic energy, 
with a lesser amount of translational kinetic energy.
We consider the Maxwell pendulum as a cylinder that is suspended
by string. The string is wound on the cylinder.

The equations of motion for Maxwell pendulum have the form
\be M_{\alpha} \frac{dv_y}{dt}=M_{\alpha}g-T , \quad 
I^{(\alpha)}_z\frac{d\omega_z}{dt}=R T , \ee
where $g$ is the acceleration such that $g\simeq 9.81 (m/s^2)$;
the axis $z$ is a cylinder axis, $T$ is a string tension,
$M_{\alpha}$ is a mass of the cylinder. 
Using $v_y=\omega_z R$, we have
\[ M_{\alpha} \frac{dv_y}{dt}=
M_{\alpha}g-\frac{I^{(\alpha)}_z}{R^2} \frac{dv_y}{dt} . \] 
As the result, we get the acceleration of the cylinder
\be \label{a-y} a^{(\alpha)}_y=\frac{dv_y}{dt}=
\frac{M_{\alpha}g}{M_{\alpha}+I^{(\alpha)}_z/R^2} . \ee
%%%=\frac{g}{1+I^{(\alpha)}_z/(MR^2)} .  \ee
Substituting Eq. (\ref{I-F}) in Eq. (\ref{a-y}), we get
\be a^{(\alpha)}_y=
%%%\frac{\alpha+2}{2\alpha+2} g
\left(1-\frac{\alpha}{2\alpha+2} \right) g . \ee
For the fractal mass dimension of the cross-section of the cylinder
$\alpha=1.5$, we get $a^{(\alpha)}_y=(3/5)g \simeq 6.87 \ (m/s^2)$.
For the cylinder with integer mass dimension of 
the cross-section ($\alpha=2$), 
we have $a^{(2)}_y=(2/3)g \simeq 6.54 \ (m/s^2)$.
The period $T^{(\alpha)}_0$ of oscillation for this Maxwell pendulum 
is defined by the equation
\[ T^{(\alpha)}_0=4t_0=4\sqrt{2L/a^{(\alpha)}_y}, \]
where $L$ is a string length, and 
the time $t_0$ satisfies the equation  $a^{(\alpha)}_yt^2_0/2=L$.
Therefore, we get the relation for the periods
\be \Bigl({T^{(\alpha)}_0}/{T^{(2)}_0}\Bigr)^2=
1+\frac{1}{3} \ \frac{\alpha-2}{\alpha+2} . \ee
%%%\be \frac{T^{(\alpha)}_0}{T^{(2)}_0}=\sqrt{{a^{(2)}_y}/{a^{(\alpha)}_y}}=
%%%\sqrt{\frac{4(\alpha+1)}{3(\alpha+2)}} . \ee

If we consider $2<D<3$ such that $1< \alpha < 2$, we can see that 
\be 8/9<\Bigl(T^{(\alpha)}_0 / T^{(2)}_0\Bigr)^2 < 1 . \ee
Note the parameter $\alpha$ can be calculated by box-counting method
for the cross-section of the cylinder.
For $\alpha=1.5$, we have $\Bigl({T^{(\alpha)}_0}/{T^{(2)}_0}\Bigr)^2=0.952$. 

A simple experiment to test the fractional continuous 
model \cite{PLA2005-1,AP2005-2} for fractal media is proposed. 
This experiment allows ua to prove that the fractional integrals 
can be used to describe fractal media.
For example, the experiment can be realized by using the sandstone.
Note that Katz and Thompson \cite{KT} presented experimental 
evidence indicating that the pore spaces of a set of sandstone samples 
are fractals and are self-similar over three to four orders of magnitude 
in length extending from 10 angstrom to 100 $\mu m$. 
The deviation $T^{(\alpha)}_0$ from $T^{(2)}_0$ is no more that 6 per cent.
Therefore, the precision of the experiments must be high.

\section{Conclusion}

In this paper we consider mechanics of fractal solids, which
are described by a fractional continuous medium model
\cite{PLA2005-1,AP2005-2}. In the general case, the fractal solid 
cannot be considered as a continuous solid.
There are points and domains that are not filled of particles.
In Ref. \cite{PLA2005-1,AP2005-2}, we suggest considering 
the fractal media as special (fractional) continuous media.
We use the procedure of replacement of the medium 
with fractal mass dimension by some continuous model that 
uses the fractional integrals.
This procedure is a fractional 
generalization of Christensen approach \cite{Chr}.
Suggested procedure leads to the fractional integration and 
differentiation to describe fractal media.
The fractional integrals are considered as approximation 
of integrals on fractals \cite{RLWQ}. Note that 
fractional integrals can be considered as integrals 
over the space with fractional dimension up to numerical factor 
\cite{Chaos2004,PRE2005,JPCS}. 
The fractional integrals are used to take into account 
the fractality of the media.

In this paper we suggest computing
the moments of inertia for fractal solids.
The simple experiments \cite{PLA2005-2} to test 
the fractional continuous model 
\cite{PLA2005-1,AP2005-2} for fractal media can be performed. 
This experiment allows us to prove that the fractional integrals 
can be used to describe fractal solid.

Note that the fractional continuous models of fractal media
can have a wide application. 
This is due in part to the relatively small numbers of parameters 
that define a random fractal medium of great complexity
and rich structure.
%%%In many cases, the real fractal structure of matter 
%%%can be disregarded and the medium can be replaced by  
%%%some fractional continuous mathematical model. 
In order to describe the media with 
non-integer mass dimension, we must use the fractional calculus.
Smoothing of the microscopic characteristics over the 
physically infinitesimal volume transform the initial 
fractal medium into fractional continuous model
that uses the fractional integrals. 
The order of fractional integral is equal 
to the fractal mass dimension of medium.
The fractional continuous model allows us
to describe dynamics for wide class fractal media 
\cite{PhysicaA,Chaos2005,AP2005-2,MPLB2005-1,POP}.

%\newpage
%%%%%%%%%%%%%%%%%%%%%%%%%%%%%%%%%%%%%%%%%%%%%%%%%%%%%%%%%%%

%%%%%%%%%%%%%%%%%%%%%%%%%%%%%%%%%%%%%%%%%%%%%%
\end{document}